\documentclass[conference]{IEEEtran}
\IEEEoverridecommandlockouts

\usepackage{cite}
\usepackage{amsmath,amssymb,amsfonts}
\usepackage{booktabs}
\usepackage{tikz}
\usepackage{amsmath}
\usepackage{listings}
\usepackage{graphicx} %
\usepackage{subcaption} %
\usepackage{rotating} %
\usepackage{svg}
\usepackage{xcolor}
\usepackage{soul, color}
\usepackage{caption}
\usepackage{array}
\usepackage[a4paper, total={184mm,239mm}]{geometry}
\def\BibTeX{{\rm B\kern-.05em{\sc i\kern-.025em b}\kern-.08em
    T\kern-.1667em\lower.7ex\hbox{E}\kern-.125emX}}
\begin{document}

\newcommand{\fuzzrducc}{FuzzRDUCC}

\title{\fuzzrducc: Fuzzing with Reconstructed\\ Def-Use Chain Coverage}

\author{
    \IEEEauthorblockN{ Kai Feng}
    \IEEEauthorblockA{
    School of Computing Science\\
        University of Glasgow \\
        Glasgow, UK\\
        k.feng.2@research.gla.ac.uk
    }
    \and
    \IEEEauthorblockN{ Jeremy Singer}
    \IEEEauthorblockA{
    School of Computing Science\\
        University of Glasgow \\
        Glasgow, UK\\
        jeremy.singer@glasgow.ac.uk
    }
    \and
    \IEEEauthorblockN{ Angelos K Marnerides}
    \IEEEauthorblockA{Dept. of Electrical \& Computer Engineering\\
       KIOS CoE\\
        University of Cyprus \\
        Nicosia, Cyprus\\
        marnerides.angelos@ucy.ac.cy
    }
}

\maketitle

\begin{abstract}



Binary-only fuzzing often struggles with achieving thorough code coverage and uncovering hidden vulnerabilities due to limited insight into a program's internal dataflows. Traditional grey-box fuzzers guide test case generation primarily using control flow edge coverage, which can overlook bugs not easily exposed through control flow analysis alone. We argue that integrating dataflow analysis into the fuzzing process can enhance its effectiveness by revealing how data propagates through the program, thereby enabling the exploration of execution paths that control flow-based methods might miss. In this context, we introduce \fuzzrducc\, a novel fuzzing framework that employs symbolic execution to reconstruct definition-use (def-use) chains directly from binary executables. \fuzzrducc\ identifies crucial dataflow paths and exposes security vulnerabilities without incurring excessive computational overhead, due to a novel heuristic algorithm that selects relevant def-use chains without affecting the thoroughness of the fuzzing process. We evaluate \fuzzrducc\ using the binutils benchmark and demonstrate that it can identify unique crashes not found by state-of-the-art fuzzers. Hence, establishing \fuzzrducc\ as a feasible solution for next generation vulnerability detection and discovery mechanisms. 



\end{abstract}


\section{Introduction}
\label{sec:intro}
Fuzzing is an effective method for detecting code vulnerabilities across diverse application domains. By generating numerous test cases and applying them repeatedly to target programs while monitoring for resulting exceptions~\cite{chen2018systematic}\cite{bohme2016coverage}, fuzzing serves as a critical technique in cybersecurity. These exceptions indicate potential security flaws. Typically, fuzzing involves a queue of seeds---interesting inputs from which new test cases are generated through mutation. These test cases are then fed into the target program to discover previously unidentified crashes.

Generally, fuzzing  focuses on maximizing control flow coverage to identify potential bugs, however this may not fully capture a program's semantics. Programs implement abstract algorithms using specific data structures and representations, and data-intensive constructs are particularly challenging to test thoroughly using only code coverage-based fuzzing \cite{gan2020greyone}.

Consider testing a parser for a structured input format like a configuration file. The parser's core logic is embedded in large data structures, with minimal code managing state transitions. Control flow coverage can confirm that parser states and transitions are executed, but it does not reflect how different inputs are processed~\cite{wang2024data}. In contrast, dataflow coverage examines how input data moves through the program, ensuring all data paths are explored and processed correctly.

This distinction is critical in fuzzing: control flow coverage might confirm that the parser runs smoothly, but only dataflow coverage can reveal whether it correctly handles a wide variety of inputs, identifying hidden bugs or vulnerabilities \cite{sofaer2024rogueone}. While control flow coverage tracks the sequence of executed operations, dataflow coverage focuses on how data (e.g., program variables) is defined, transformed, and used during runtime. It traces the interactions between where data is defined and where it is used, referred to as definition-use (def-use) chains. By analyzing both control and data flow, fuzzing can more thoroughly test complex, data-driven program constructs.

Recent studies like Greyone~\cite{gan2020greyone} and DataFlow~\cite{herrera2023dataflow} present lightweight instrumentation that leverages def-use information to assist fuzzers in generating test cases. However, these methodologies are tightly coupled to the LLVM compiler and require access to source code, limiting their applicability. This constraint poses challenges in analyzing binary files, such as firmware commonly found in IoT environments, where dataflow-based fuzzing solutions are lacking~\cite{yun2022fuzzing}. Additionally, these methods require more resources than control flow coverage, as they depend on acquiring preliminary information and tracking dynamic updates during execution.

Therefore, this paper is structured around three pivotal research questions:

\begin{itemize} 
\item \textbf{RQ1}: Is it feasible to implement coverage-based fuzzing by leveraging the inherent dataflow within binaries? 
\item \textbf{RQ2}: How does the runtime overhead of dataflow-based fuzzing compare to that of control flow fuzzing? 
\item \textbf{RQ3}: Does dataflow coverage-based fuzzing approach identify a broader or alternative set of bugs in standard benchmarks compared to control flow fuzzing? 
\end{itemize}

To address these questions, we introduce a novel methodology that emphasizes dataflow tracking within binaries without debug symbols, moving away from the traditional focus on control flow and source code. We use the Angr~\cite{shoshitaishvili2016state} symbolic execution framework to extract and select def-use chains from binary code according to a heuristic algorithm. By integrating dataflow analysis directly into execution, our approach provides precise feedback to the fuzzer.

The contributions of \fuzzrducc\ framework are as follows:

\begin{enumerate}
\item \textbf{Recovery of def-use chains}: Our method reconstructs def-use chains directly from binary code, enabling thorough static analysis.
\item \textbf{Efficient dataflow tracking with lightweight instrumentation}: We track def-use chain coverage via program counter hashing during dynamic binary rewriting.
\item \textbf{Binary-focused dataflow scheme}: We analyze the dataflow graph directly from the binary, providing a comprehensive understanding of runtime execution paths to enhance fuzzing performance.
\item \textbf{Heuristic selection of def-use chains}: To minimize runtime overhead, we propose a heuristic algorithm that efficiently selects appropriate defs and uses for instrumentation.
\end{enumerate}

\section{Technical Background}
\label{sec:technical_background}
\subsection{The fuzzing for binary}

American Fuzzy Lop (AFL) \cite{zalewski2017american} and AFL++ \cite{fioraldi2020afl++} have gained widespread recognition within the research community as a quality baseline for fuzzing research. Numerous studies have developed their methodologies based on AFL's capabilities. AFL is a grey-box fuzzer that generates test cases using a variety of mutation strategies tailored to achieve comprehensive code coverage. For binary fuzzing, AFL incorporates QEMU \cite{bellard2005qemu}, a generic and open-source machine emulator and virtualizer, to emulate the execution of binaries. This emulation facilitates the addition of instrumentation, enabling AFL to obtain feedback from the binary's execution to obtain the binary's control flow. When new code coverage is discovered, AFL adapts its mutation strategy based on the test case associated with this coverage.

\subsection{Towards dataflow coverage}
\lstset{
  language=C,  
  basicstyle=\footnotesize\ttfamily, 
  frame=shadowbox, 
  rulesepcolor=\color{gray!30}, 
  keywordstyle=\color{blue!90}\bfseries, 
  commentstyle=\color{green!60}\textit, 
  showstringspaces=false, 
  numbers=left, 
  numberstyle=\tiny\color{gray}, 
  stringstyle=\color{orange}, 
  breaklines=true, 
  breakatwhitespace=true, 
  tabsize=4, 
  captionpos=b, 
  xleftmargin=10pt, 
  xrightmargin=10pt, 
  morekeywords={protocol, imffield}, 
  extendedchars=true 
}

\begin{lstlisting}[caption={Fuzzing Code Example},label=uivulnerability]
/* If the first 4 bytes are 0x01f401f4 (udp src and dst port = 500) we most likely have UDP (isakmp) traffic */
if (tvb_get_ntohl(tvb, 0) == 0x01f401f4) { 
        protocol = TCP_ENCAP_P_UDP;
    } else { 
        protocol = TCP_ENCAP_P_ESP;
}
if (g_ascii_strcasecmp(header_name, "Content-Length") == 0) {
} else (g_ascii_strcasecmp(header_name, "Transfer-Encoding") == 0) {
/* Process Transfer-Encoding header and other headers */
}

\end{lstlisting}

While code coverage is a powerful tool in fuzzing, it has shortcomings when dealing with data-intensive program constructs. In Listing~\ref{uivulnerability}, an \texttt{if} statement from Wireshark checks whether the first four bytes of a packet match the specific magic number \texttt{0x01df401f4}, indicating UDP traffic (specifically ISAKMP). Code coverage can only indicate if this condition is true or false. However, when the condition fails, code coverage does not reflect how close the input is to the target value. Without proper guidance, the fuzzer must blindly guess the correct value, facing a probability of success of $1$ in $2^{32}$\cite{wang2024data}.

Two common strategies to address such branches are concolic execution~\cite{yun2018qsym}\cite{borzacchiello2021fuzzolic} and intelligent branch solving~\cite{mallissery2023demystify}\cite{yi2024compatible}\cite{chen2018angora}.

Concolic execution models constraints as symbolic expressions, allowing solvers like SMT solvers to find solutions\cite{felli2023data}. By treating input bytes as sequences of 8-bit vectors, we update their symbolic representations during execution. However, constraint solvers often struggle with simple string comparisons. Although concolic execution can systematically explore program paths to solve these constraints, it cannot differentiate between meaningful and superficial path differences. For instance, the function \texttt{g\_ascii\_strcasecmp} performs case-insensitive comparisons. Different headers like \texttt{Content-Length} and \texttt{Transfer-Encoding} result in distinct paths, even when header order changes, leading to path explosion and resource exhaustion.

Intelligent branch solving struggles with this issue without manual intervention. Since comparing a single character can easily succeed, the branch is quickly marked as solved, and further analysis is skipped. As a result, constraints for the remaining characters never reach the solver. Modern fuzzers attempt to mitigate this with program-specific optimizations. For example, AFL++ uses \textit{CmpLog} instrumentation to record operands of failed comparisons and applies heuristics to solve them \cite{aschermann2019redqueen}. Instead of instrumenting branches in a general way, it relies on a hard-coded list of comparison functions, treating each call as an abstract branch-a clearly non-scalable approach.

To improve these methods, dataflow coverage provides a more precise approximation of program behavior by focusing on how variables are assigned and used, rather than just the sequence of executed operations. This approach considers more complex structures instead of solving constraints, like lookup tables, binary trees, and directed graphs, offering deeper insights into program execution. By shifting from a control flow to a dataflow perspective, fuzzing techniques can be made more effective~\cite{herrera2023dataflow}.

\section{\fuzzrducc\  Framework}
\label{sec:methodology}
\begin{figure}
    \centering
    \includegraphics[width=0.5\textwidth]{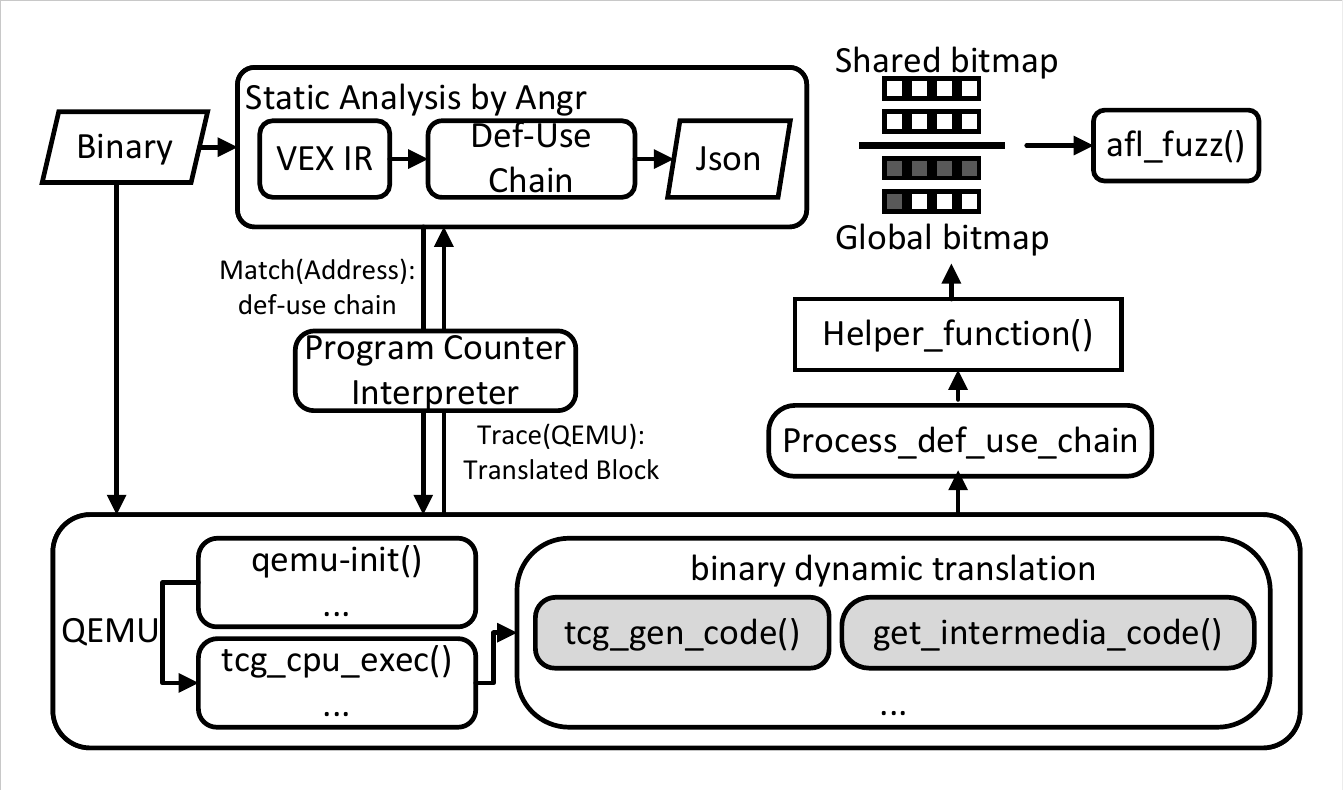}
    \caption{Structure of \fuzzrducc}
    \label{fig:enter-label}
\end{figure}
Our approach enhances fuzzer effectiveness by incorporating def-use chains, structured into two main phases: static analysis and fuzzing (see Figure~\ref{fig:enter-label}). In the static analysis phase, we use the Angr framework to extract def-use chains from the binary, obtaining precise addresses and counts of defs and uses for each translated block. This involves instrumenting the code to record the addresses and numbers of defs and uses, leveraging QEMU's lightweight code generation.

In the fuzzing phase, we repurpose the AFL++ bitmap (previously used as a proxy for edge coverage) to monitor the coverage of def-use chains accurately. 
As each basic block executes, we update the local AFL++ bitmap against a global map to track changes in execution state.
This mechanism guides the fuzzer to re-mutate inputs based on analysis of previous seeds, aiming to significantly improve fuzzing efficiency by combining static analysis with dynamic fuzzing.

\section{Methodlogy and Implementation}

\subsection{Def-Use Chain Generation}

We extract def-use chains from binaries through symbolic execution using the Angr framework, departing from traditional methods that rely on Static Single Assignment (SSA) form. Angr loads the binary components, including library dependencies, and uses  VEX Intermediate Representation (IR) to reconstruct the control flow graph and dataflow graph directly from machine code. This process maps out the program's execution flow and provides a representation of all possible execution paths, enabling comprehensive analysis of the binary's execution semantics~\cite{kim2017testing}.

Our method incorporates reaching definition analysis~\cite{reps1995precise} to determine where variables (definitions) are assigned values and where these values are used across different basic blocks. This analysis reveals relationships between definitions and uses in the code, identifying uses reachable from definitions that have not been overwritten, using an over-approximation strategy. While this may sacrifice some soundness, it offers increased speed in analyzing binaries, which we consider acceptable for achieving sufficient precision in binary-level analysis.

By storing the def-use chains in a JSON file, we facilitate their integration into QEMU's code generation process during dynamic binary translation. By systematically identifying and analyzing def-use chains in binaries, we lay the groundwork for more effective fuzzing strategies by enhancing our ability to uncover vulnerabilities through understanding dataflow. This methodology outlines potential pathways through which definitions affect uses, providing a solid foundation for tracking interrelations and dependencies within the code. By enabling a focused exploration of the software's execution space, it enhances the precision and efficiency of fuzzing processes, thereby improving vulnerability detection through a thorough understanding of the software's internal mechanisms.

\subsection{Code Instrumentation}

After reconstructing the def-use chains, we integrate them into the execution of the binary managed by QEMU, which decomposes the binary into basic blocks. Each block is translated into a host-specific block through QEMU's Tiny Code Generator (TCG), converting each instruction into micro-operations within the translated block. During dynamic binary translation, these instructions are transformed into host instructions tailored to the specific architecture.

We adapt QEMU to utilize its tracing capabilities to obtain information about translated blocks, specifically retrieving the Program Counter (PC) value for each executed block. The TCG functions as a just-in-time compiler, translating guest instructions into executable code for the host architecture. By retaining the guest PC for each block and employing a hash table to associate it with the host PC of the translated block, we achieve precise tracking of control flow and dataflow.

The translation process in QEMU is divided into a frontend and a backend. The frontend lifts target instructions into TCG Intermediate Representation (IR), which is stored in a list. We focus on tracing the current execution of translated blocks (TBs), particularly utilizing the cache list to identify the current TB and obtain its PC. We then correlate the PC with the def-use chains stored in the JSON file generated by Angr, mapping the def-use chain within each translated block.

With the def-use information for the binary, we apply precise instrumentation to monitor identified definitions and uses within these blocks. After acquiring the def-use chain for each translated block, the backend converts the TCG IR into host machine code. TCG IR registers are categorized into various types: global, local temporary, normal temporary, fixed, constant, and extended basic block (ebb). Our objective is to encapsulate definitions and uses within TCG registers, generating corresponding IR to embed into the translated block for recording purposes.

We utilize helper functions to pass parameters to registers and execute jumps to specific addresses. These helper functions can also access the CPU environment, enhancing our ability to manipulate and track the execution flow.

This approach mirrors the \texttt{afl\_maybe\_log} function used in AFL++, which inserts IR into the translated block to monitor execution. However, our instrumentation focuses on usages rather than recording every definition and usage, recognizing that in statically compiled binaries, some definitions may not be utilized or analyzed correctly. Focusing on usages is critical for understanding data manipulation.

By integrating def-use chain information with QEMU's execution trace, we gain deeper insights into execution patterns, facilitating more targeted fuzzing to uncover vulnerabilities. This dynamic tracking of data and control flow enables precise identification and analysis of critical execution paths and enhances our ability to detect and assess the impact of definitions and uses throughout the software's operation.

\subsection{Optimizing Def-Use Chain Selection}

Instrumenting all def-use chains in translated blocks introduces significant time and space overhead during fuzzing. For example, in the binutils dataset, one binary's translated block size increased fivefold after instrumentation~\cite{herrera2023dataflow}. To mitigate this overhead, we propose a heuristic algorithm that selectively targets addresses of common external library functions and optimizes the def-use chain selection process.

To reduce overhead, we exclude definitions and uses within the same block or function, thereby reducing the size of the translated blocks. We also disregard definitions that are not used or not detected by Angr, streamlining the analysis process. By calculating the distance between definitions and their related uses, we focus on definitions and uses that span across different functions, utilizing interprocedural analysis to efficiently identify the necessary def-use chains.

Angr's simulation involves instruction emulation and symbolic execution for branch decisions, maintaining stacks of states with register values and memory addresses. State duplication can lead to explosion, especially in loops dependent on user input, causing delays in vulnerability detection. To optimize analysis, we use Angr to identify addresses of common libc functions such as \texttt{malloc}, \texttt{calloc}, and \texttt{free}, avoiding detailed examination of external library functions. We employ Angr's SimProcedures to replace third-party library functions with custom implementations that simulate behavior, which is important for statically compiled targets where analyzing external libraries is resource-intensive.

We designed custom hook functions (\texttt{handle\_malloc}, \texttt{handle\_calloc}, \texttt{handle\_free}) for SimProcedures to simulate memory management effects on the analysis state. For example, if \texttt{malloc} is at address \texttt{0x400900} in the binary, Angr hooks this address with its SimProcedure for \texttt{malloc}. When execution reaches \texttt{0x400900}, the SimProcedure is invoked instead of the actual \texttt{malloc}, allowing efficient reaching definition analysis on specific addresses while focusing resources on primary binary analysis.

This heuristic captures ``interesting'' def-use chains, increasing the likelihood of discovering new crashes and exploring more code paths. It enhances the efficiency of our instrumentation and improves the overall effectiveness of our fuzzing strategy.

Our approach adapts AFL++ to track dataflow rather than control flow. After identifying def-use chains within a target binary, we introduce an alternative coverage bitmap to track changes in these def-use chains. This bitmap records runtime relationships between definitions and uses, logging any changes observed.

When a modification in this bitmap indicates a change in dataflow coverage, we initiate a strategic re-mutation of the seed. This re-mutation aims to explore unexplored code paths, broadening coverage and deepening the fuzzing process. This method ensures a nuanced and dynamic examination of the binary's behavior, enhancing the potential for identifying vulnerabilities.

\subsection{Updating the Coverage Scheme}

AFL++ uses QEMU's TCG IR to insert instrumentation code that computes a hash for each edge during execution. An ``edge'' represents a transition between code blocks (e.g., from block $A$ to block $B$). For each transition, AFL++ generates a unique identifier $i$ by hashing the addresses of both source ($A$) and destination ($B$) blocks, with the source address right-shifted: \begin{equation} i \gets \mathrm{addressof}(B) \oplus (\mathrm{addressof}(A) \gg 1). \end{equation} This hash $i$ indexes into the edge coverage bitmap, where each index represents a potential execution edge. When an edge is traversed, AFL++ increments the value at that index, enabling it to monitor executed edges and prioritize inputs that explore new paths.

We adapt AFL++'s tracking to capture the relationship between definitions and uses. At code generation time, we use precomputed def-use chains (from Angr's JSON files) for each translated block and employ a helper function to embed def-use instrumentation into the TCG IR.

Since every block has definitions, inserting IR into every block can slow translation. Therefore, we focus on usages, tracing them to identify related definitions. We document def-use edges according to the def-use chains. Our revised hash function for the coverage bitmap index uses the addresses of the definition and use sites: \begin{equation} i \gets \mathrm{addressof}(\mathit{def}) \oplus \mathrm{addressof}(\mathit{use}). \end{equation} Using these addresses as hash values generates a unique $i$, reducing collision risk. This approach allows tracing multiple definitions and uses within a single block, providing nuanced and sensitive coverage feedback. It enhances analysis granularity and improves fuzzing efficiency by focusing on critical dataflow aspects of program execution.

\section{Evaluation}
\label{sec:evaluation}

Our primary objective is to address the three research questions outlined in Section~\ref{sec:intro}, focusing on the feasibility and efficiency of dataflow coverage-based fuzzing for software testing. We aim to compare it with control flow-based fuzzing and evaluate its potential to uncover unique vulnerabilities.


\begin{figure*}[!th]
    \centering
    \begin{minipage}{\textwidth} 
        \centering
        \includegraphics[width=0.4\linewidth]{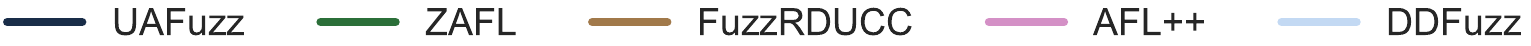}
    \end{minipage}


    \begin{minipage}{\textwidth}
        \centering
        \includegraphics[width=0.24\textwidth]{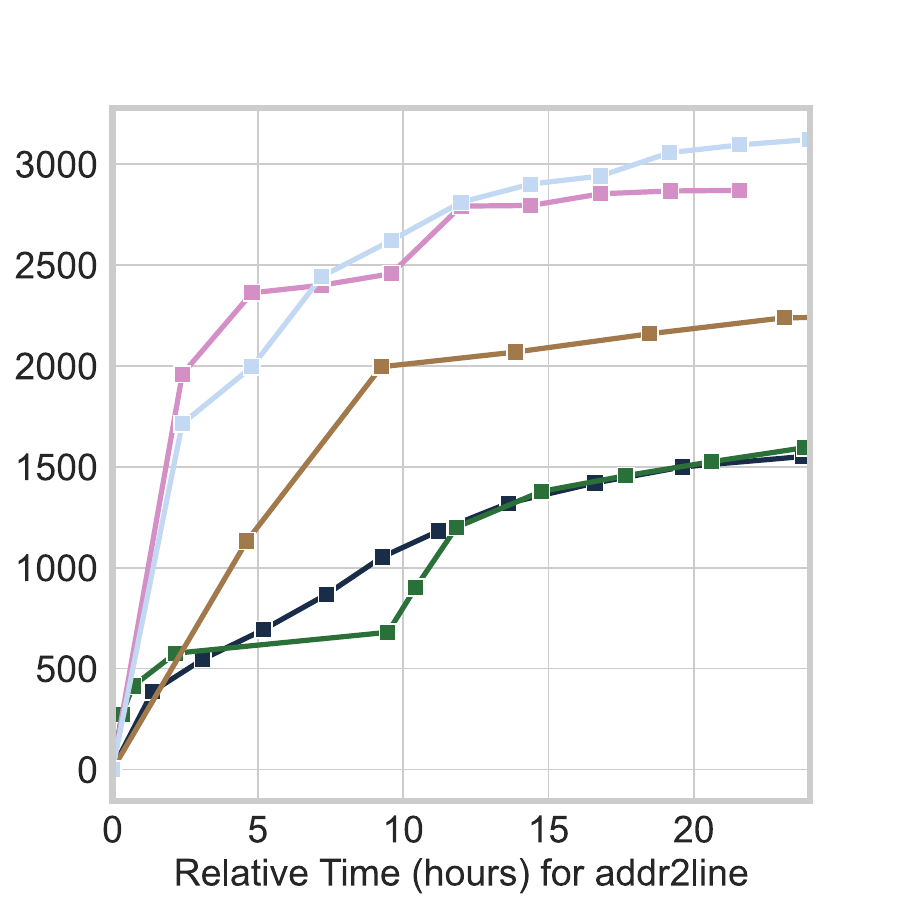}\hfill
        \includegraphics[width=0.24\textwidth]{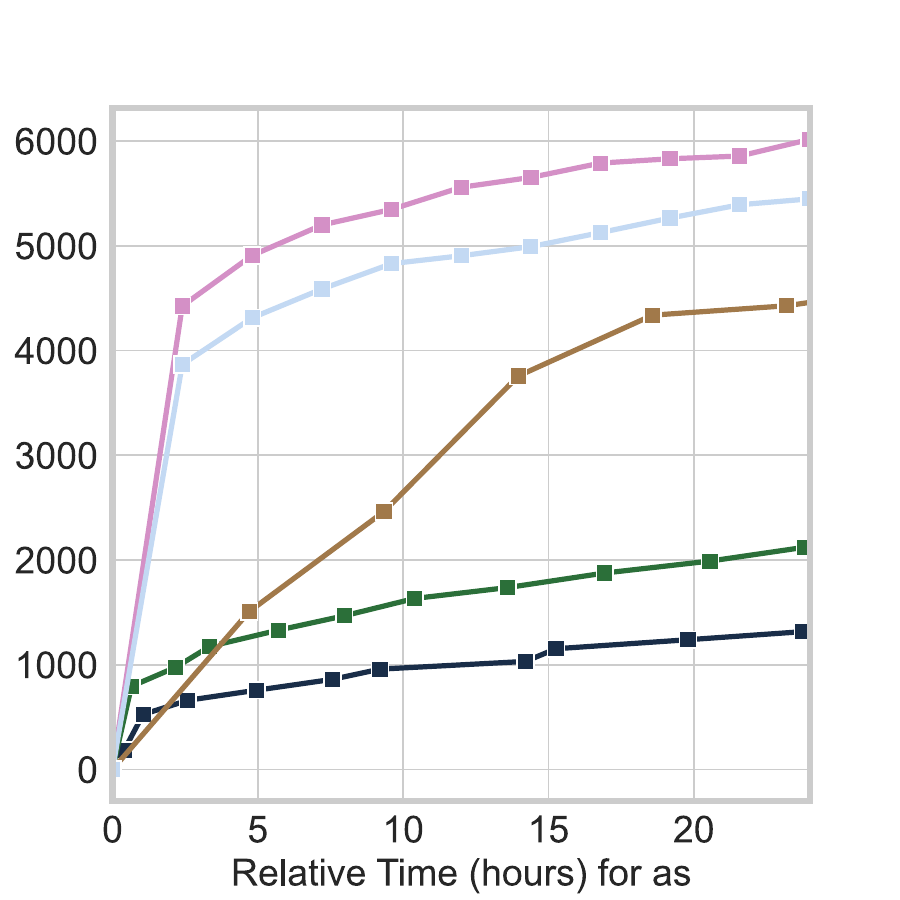}\hfill
        \includegraphics[width=0.24\textwidth]{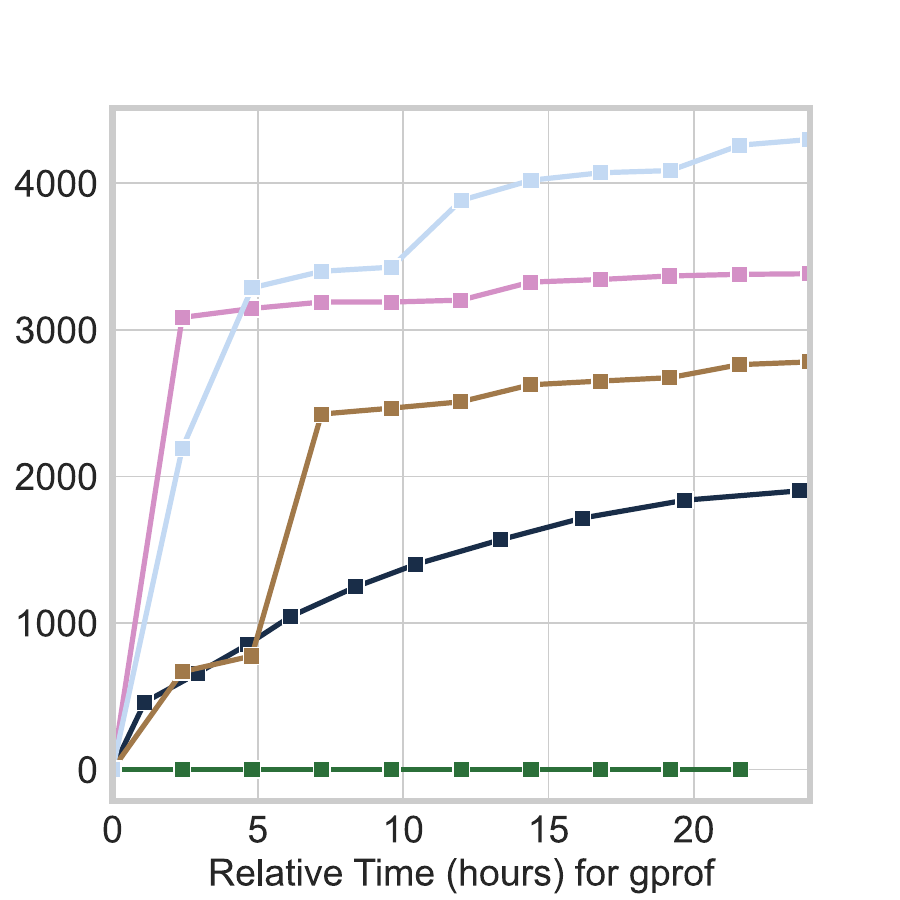}\hfill
        \includegraphics[width=0.24\textwidth]{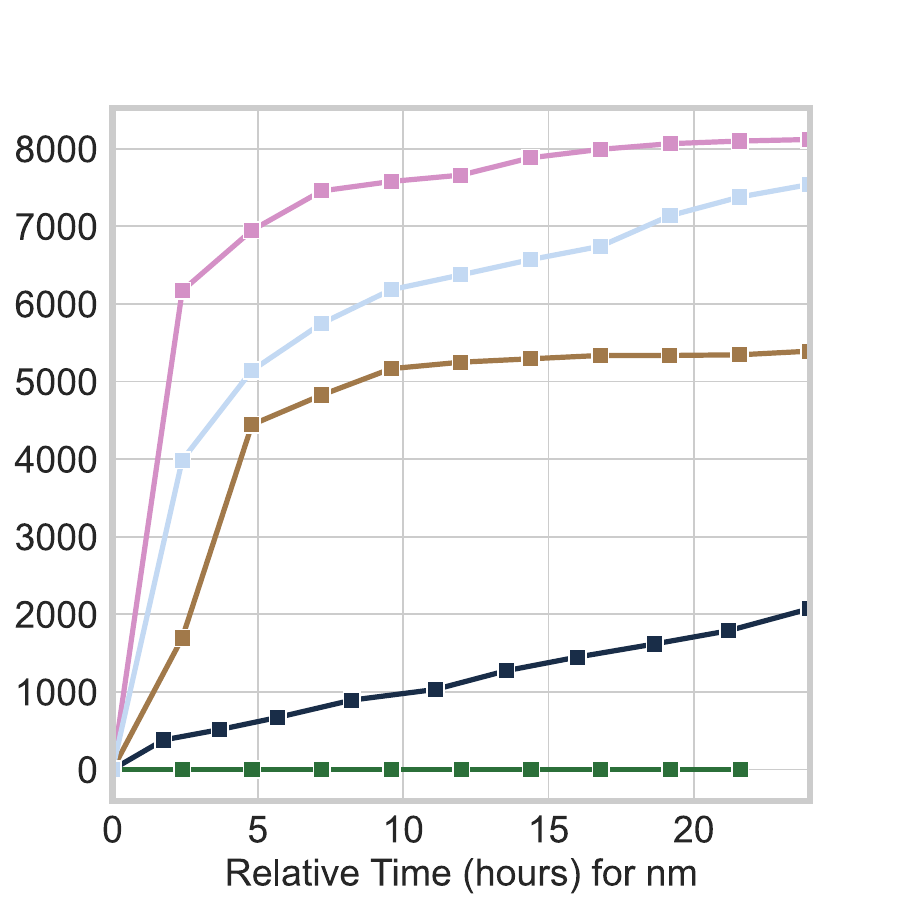}
    \end{minipage}
    
    \vspace{-0.1cm} 

    \begin{minipage}{\textwidth}
        \centering
        \includegraphics[width=0.24\textwidth]{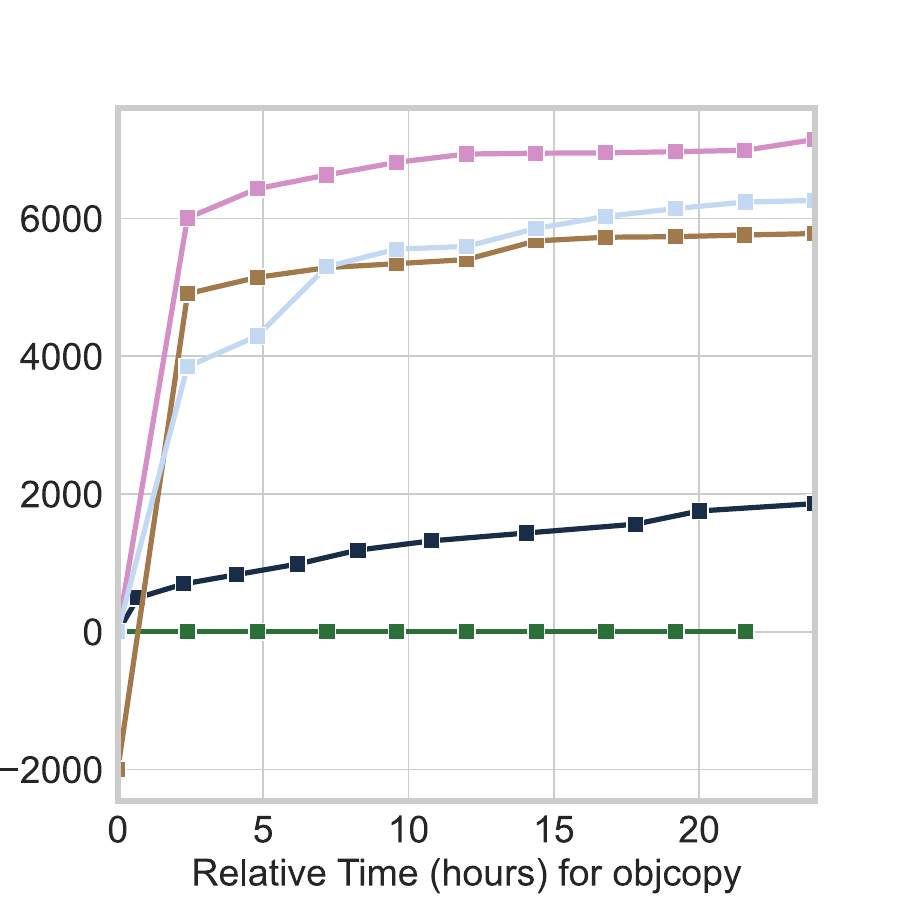}\hfill
        \includegraphics[width=0.24\textwidth]{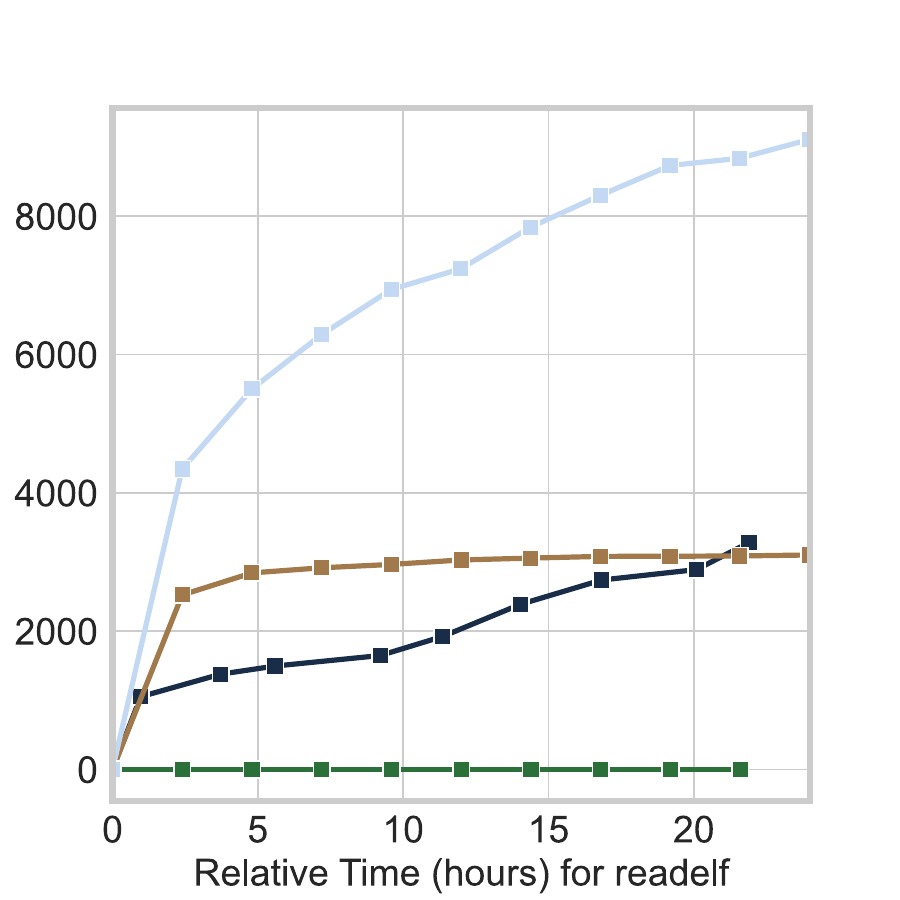}\hfill
        \includegraphics[width=0.24\textwidth]{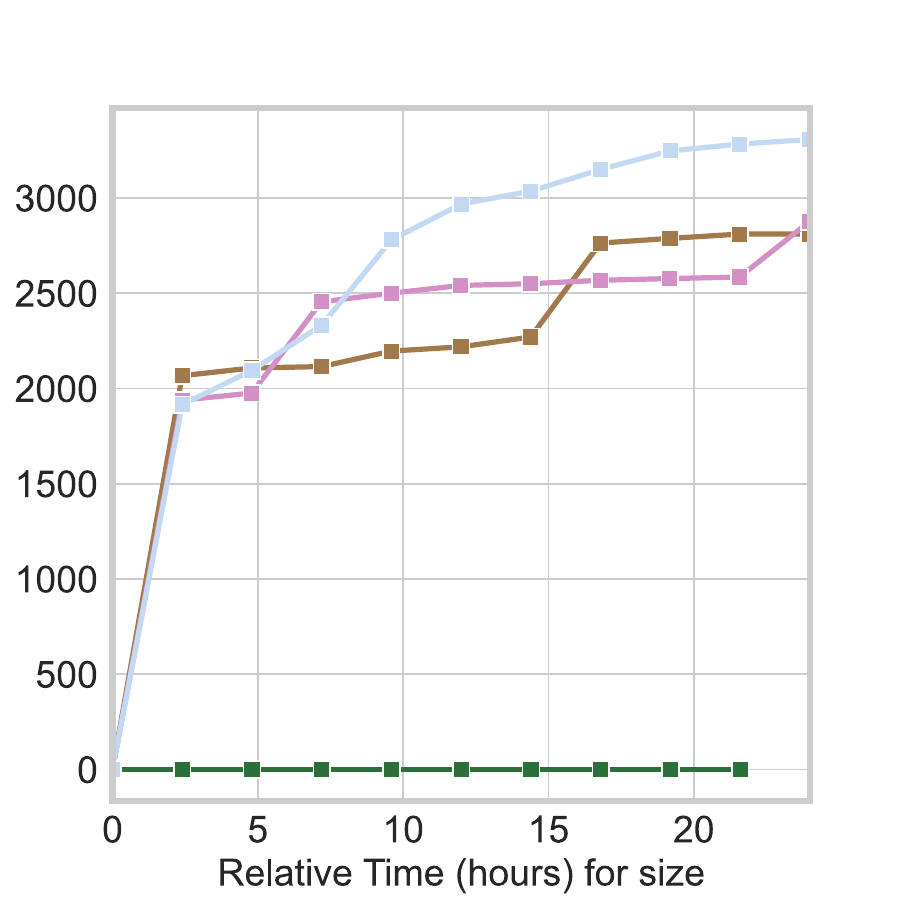}\hfill
        \includegraphics[width=0.24\textwidth]{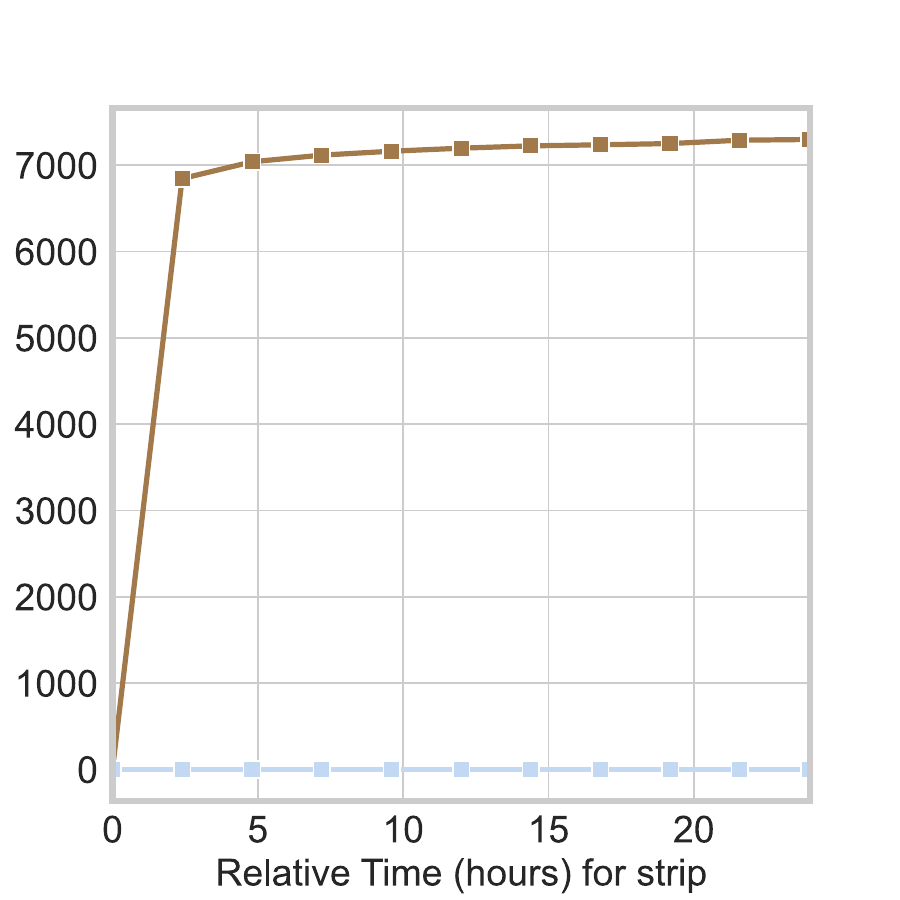}
    \end{minipage}
    
    \caption{Change of Edge Coverage ($y$-axis) within 24 hours ($x$-axis) for Fuzzing of Binutils Tools}
    \label{fig:edge}
\end{figure*}

\subsection{Evaluation Setup}

\textbf{\textit{Baseline}}: In this study, we compare our proposed framework with several established fuzzing tools, including AFL++, DDfuzz\cite{mantovani2022fuzzing}, uafuzz \cite{nguyen2020binary}, and ZAFL \cite{nagy2021breaking}. Our analysis focuses on their design principles and operational efficiencies, particularly in relation to dataflow analysis. AFL++ serves as the baseline for fuzzing comparisons, building upon features from AFLfast, Mopt \cite{lyu2019mopt}, and other advancements in AFL-based tools. DDfuzz introduces a dataflow-driven feedback mechanism that extends beyond control flow edge discovery by guiding fuzzing based on a data dependency graph, although it only supports source code. uafuzz specializes in binary-directed fuzzing to detect use-after-free vulnerabilities, using novel seed metrics to select appropriate seeds for mutation. ZAFL, on the other hand, enhances binary-only fuzzing through binary rewriting to achieve compiler-quality instrumentation. Notably, except for ZAFL, all the mentioned fuzzers utilize the QEMU model, a common framework for emulation-based fuzzing.

\textbf{\textit{Dataset}}: To comprehensively assess the effectiveness of the fuzzing frameworks, we use a diverse set of binaries, with a focus on GNU Binutils—a widely used suite of binary tools. The selection of Binutils is driven by its critical role within the Linux ecosystem, making it a well-established benchmark for fuzzing evaluations. We test 8 binaries from the Binutils collection to evaluate the fuzzers under consideration.


    



\textit{\textbf{Evaluation Metric}}

\begin{itemize}
    \item \textbf{The change of edge Coverage}:  The effectiveness of a fuzzer is measured by its ability to increase edge coverage, which provides valuable insights into the program's execution paths. Edge coverage acts as a feedback mechanism, helping the fuzzer explore uncharted execution paths and uncover potential vulnerabilities. This metric is pivotal in gauging how thoroughly a fuzzer explores the program's execution space.

    \item \textbf{Number of Crashes}:  Crashes are a key indicator of a fuzzer's efficacy, as they signal the discovery of potential vulnerabilities or software defects. A higher number of crashes directly correlates with the fuzzer’s ability to uncover significant issues in the target software, making crash detection a vital evaluation criterion.
    
    \item \textbf{Execution Speed}: When fuzzing binary targets, particularly with QEMU-based fuzzers, there are inevitable performance overheads. It is important to balance between maximizing edge coverage and minimizing performance degradation. Assessing this trade-off is crucial to determine a fuzzer’s effectiveness in binary analysis, as maintaining efficiency while reducing overhead is key to successful fuzzing operations.

\end{itemize}

\subsection{Preliminary Results}

\subsubsection*{RQ1: Assessing the Feasibility of Implementing Coverage-Based Fuzzing through Dataflow Analysis}

To address the first question (Q1), we analyze changes in edge coverage as depicted in Figure \ref{fig:edge}. Higher change of edge coverage indicates the effectiveness of the fuzzer's feedback mechanism in uncovering more paths within the binary. AFL++ and DDfuzz show the best performance in edge detection among the fuzzers evaluated. \fuzzrducc\  surpasses UAFuzz and ZAFL in coverage metrics within a 24-hour testing period. This enhanced performance results from the implementation of a novel instrumentation method that meticulously instruments both uses and corresponding defs. As dataflows through these instrumented edges, they are dynamically updated, highlighting the benefits of incorporating dataflow information beyond traditional control flow analysis. Notably, in the strip benchmark, edge coverage remains unchanged among other fuzzers, except for our framework, which demonstrates distinct improvement. 

\begin{figure}
    \centering
    \includegraphics[width=0.4\textwidth]{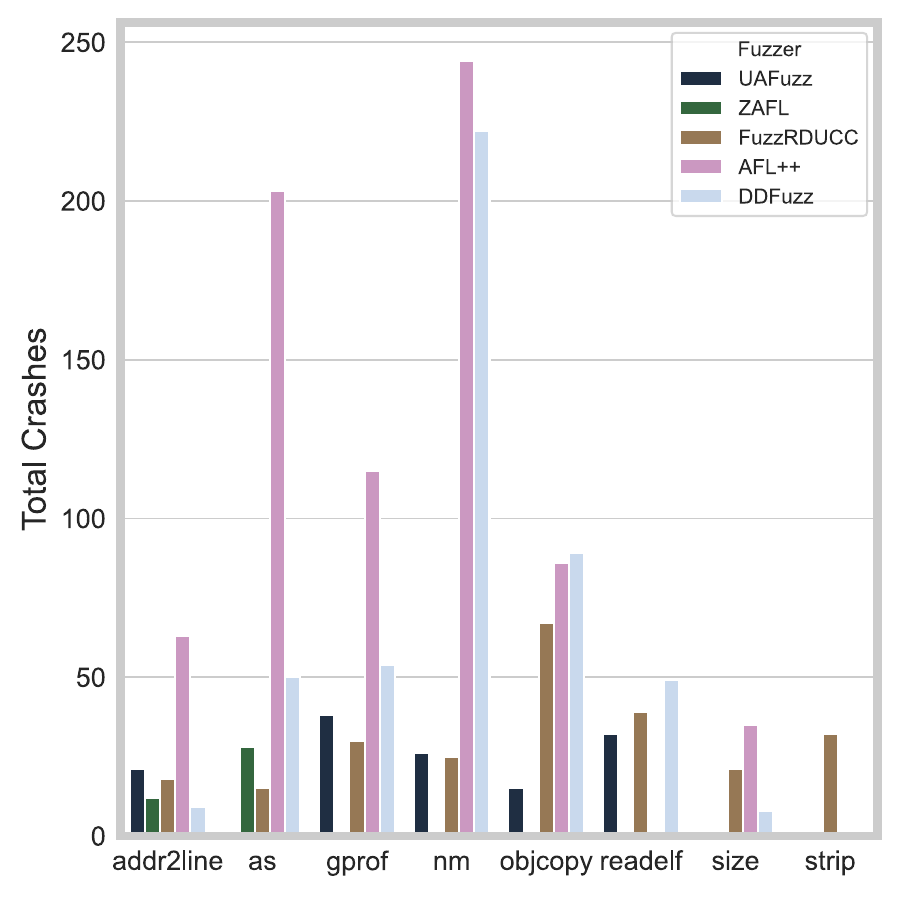}
    \caption{Comparison of Crashes Across Different Targets for Each Fuzzer}
    \label{fig:crash}
    \vspace{-0.5cm}

\end{figure}
\begin{figure}[t]
    \centering
    \includegraphics[width=0.4\textwidth]{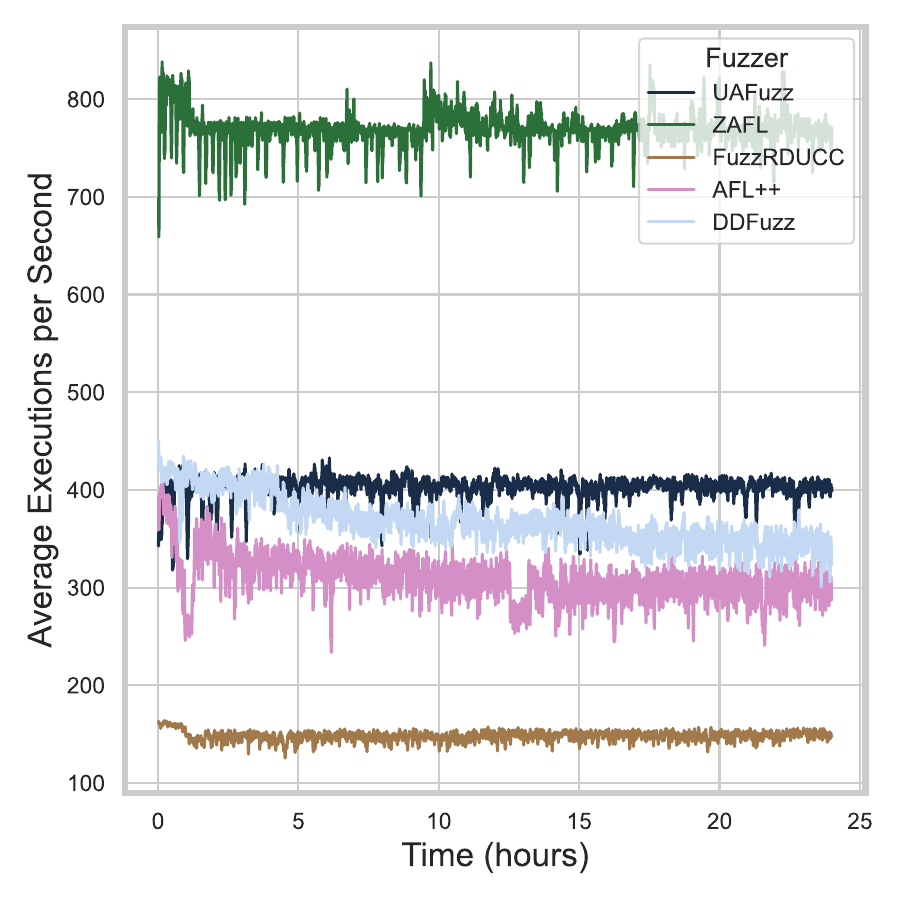}
    \caption{Average Execution Speed Over Time for Each Fuzzer}
    \label{fig:execution}
\end{figure}
\subsubsection*{RQ2: Benchmarking the Bug Discovery Capabilities of Dataflow Coverage-Based Fuzzing}

Figure \ref{fig:crash} illustrates the bug detection rate over a 24-hour period. Although our framework achieves higher edge coverage, this does not necessarily correlate with a higher rate of crash discoveries in binaries. AFL++ demonstrates the best performance across all benchmarks. Notably, our framework identified unique crashes within the strip benchmark that are not detected by other fuzzers, highlighting its ability to uncover distinct vulnerabilities.

This def-use chain analysis allows the fuzzer to generate inputs that not only conform to the expected binary format but also manipulate specific internal structures. In the case of strip, FuzzRDUCC generates inputs that influence the processing of relocation entries, guiding execution toward deeper code regions where vulnerabilities, such as the null pointer dereference in \texttt{copy\_relocations\_in\_section}, reside.

Specifically, the crashes are locate within the \texttt{copy\_relocations\_in\_section} function when the program attempts to dereference a null or invalid pointer in a loop processing relocation entries. The null pointers leads to a segmentation fault when processing malformed inputs.

In contrast, other fuzzers like AFL++ rely primarily on control flow-based coverage metrics and employ generic, format-agnostic mutations. These approaches produce inputs that are rejected early in the execution process due to failing initial format validations, preventing them from exploring the deeper code paths where the vulnerability exists. Consequently, they neither achieve significant edge coverage in strip nor discover the associated crashes.

\subsubsection*{RQ3: the Runtime Overhead for Dataflow-Based Fuzzing Compared to Control Flow Fuzzing}

Figure \ref{fig:execution} illustrates the average execution time across all datasets for different fuzzers. Despite implementing the selection algorithm for the def-use chain, significant overhead persists in analyzing the dataflow. However, there is a noticeable improvement in performance with the introduction of the selection algorithm. Previously, the execution rate was approximately 50 executions per second; the implementation of this algorithm has substantially improved this to 150 executions per second.

\subsection{Future Evaluation} 

Building on our preliminary evaluation, we plan to conduct the following in-depth experiments:

\subsubsection*{1. Reducing Overhead via Selective Def-Use Chain Implementation}

We aim to significantly reduce binary analysis overhead by selectively implementing def-use chains. By analyzing and categorizing each def-use chain using various static analysis algorithms, we will identify the most impactful chains affecting memory changes, striving for a balance between soundness and completeness. We will also simplify def-use chains related to heap memory and optimize the size of translated blocks in QEMU to further decrease emulation overhead.

\subsubsection*{2. Enhancing Fuzzing Performance with Def-Use Chain Guidance}

We hypothesize that def-use chain-guided fuzzing will outperform traditional methods in triggering crashes and detecting vulnerabilities. By focusing fuzzing efforts on specific def-use chains that represent critical paths potentially harboring vulnerabilities, we aim to increase efficiency and uncover flaws that broader methods may miss. This approach also involves identifying unique vulnerabilities among the detected crashes.

\subsubsection*{3. Developing Dataflow Coverage Metrics}

To better understand the impact of def-use chains, we plan to propose a new metric for dataflow coverage. Since existing fuzzers primarily use bitmaps to track control flow coverage—insufficient for representing dataflow trends—introducing a dataflow coverage metric will help us assess coverage more accurately. This metric will also assist in selecting appropriate hash functions for computing def-use chains, thereby reducing hash collisions.

\subsubsection*{4. Applying the Framework to Real-World Scenarios}

To validate our hypotheses, we are conducting baseline evaluations with GNU Binutils. We will extend our tests to other datasets, such as Magma~\cite{hazimeh2020magma} and Fuzzbench~\cite{metzman2021fuzzbench}, and plan to experiment with real-world IoT firmware using datasets from~\cite{feng2020p2im,clements2020halucinator}. These experiments will demonstrate our framework's adaptability and effectiveness across diverse environments.

\section{Conclusion}
\label{sec:conclusion}
We have established the feasibility of using reconstructed def-use chains as feedback to drive the fuzzing process. We have developed a framework designed to recover def-use chains from binary code, thereby providing a new coverage mechanism for grey-box fuzzers. Preliminary results on binutils suggest that our framework successfully identifies some unique crashes, although it incurs relatively high overhead. Future work will focus on reducing this runtime overhead and conducting more comprehensive evaluations. We also provide source code required to replicate the experiments presented in this paper.\footnote{https://github.com/MaksimFeng/AFLplusplus}

\bibliographystyle{unsrt}
\bibliography{reference}

\end{document}